\documentclass[a4paper]{article}
\usepackage{graphicx}
\usepackage{amsmath}

\begin{document}

\begin{center}
{\bf \Large Average distance in growing trees}\\[5mm]

{\large K.Malarz$^*$, J.Czaplicki, B.Kawecka-Magiera and K.Ku{\l}akowski$^\dag$}

\bigskip

{\em Department of Applied Computer Science,
Faculty of Physics and Nuclear Techniques,
AGH University of Science and Technology\\
al. Mickiewicza 30, PL-30059 Krak\'ow, Poland\\

\bigskip

E-mail: $^*$malarz@agh.edu.pl,
$^\dag$kulakowski@novell.ftj.agh.edu.pl\\[3mm]

April 28, 2003
}
\end{center}

\begin{abstract}
\noindent
Two kinds of evolving trees are considered here: the exponential trees, where
subsequent nodes are linked to old nodes without any preference, and the
Barab\'asi--Albert scale-free networks, where the probability of linking to a
node is proportional to the number of its pre-existing links. In both cases,
new nodes are linked to $m=1$ nodes.  Average node-node distance $d$ is
calculated numerically in evolving trees as dependent on the number of nodes
$N$. The results for $N$ not less than a thousand are averaged over a thousand
of growing trees. The results on the mean node-node distance $d$ for large $N$
can be approximated by $d=2\ln(N)+c_1$ for the exponential trees, and
$d=\ln(N)+c_2$ for the scale-free trees, where the $c_i$ are constant. We
derive also iterative equations for $d$ and its dispersion for the exponential
trees. The simulation and the analytical approach give the same results.
\end{abstract}


\section{Introduction}

In recent five years, much attention has been paid to the problem of evolving
networks \cite{ab1,drm,nwm}. For our purposes, the problem can be
summarized as follows: we consider a graph, initially small, of $N$ nodes.
A new $(N+1)$-th node is linked to $m$ nodes, selected randomly from $N$
previously existing nodes, with probability of linking to a given node
dependent on its degree, i.e. its number of edges $k$. If this probability is
constant, we get the so-called exponential network --- this term is due to the
exponential distribution $P(k)$ of the number of edges of a node \cite{drm}.
The distribution $P(k)$ is often treated as giving most essential local
characteristics of the network. Example giving, for the Cayley tree
$P(k)=\delta_{k,3}$.  If the probability of linking to a node is proportional
to the node degree $k$, the network is scale-free, i.e. the distribution
$P(k)\propto k^{-\gamma }$, where $\gamma \approx 3.0$. This kind of networks
is called ``Barab\'asi--Albert networks'' from the names of the inventors
\cite{r1}. The scale-free character has been discovered in many real networks:
a network of web pages which are available one from another, a network of
coauthors of scientific papers, a network of actors playing roles in the same
movies and some others. In any case, the evolving procedure is to be
considered as a non-equilibrium process.

Here we focus on the mean distance $d$ between nodes in trees, where
$m=1$. In trees, the path between each two nodes is unique and it cannot be
changed during the system evolution. A number of results on $d$ in trees
has been derived recently for uncorrelated random networks
\cite{burda,dms,hol2,bia}. These results are based on scaling hypotheses and/or
assumptions of lack of correlations between nodes. The goal of the present work
is to calculate $d(N)$ numerically for the exponential trees and the
Barab\'asi--Albert $m=1$ scale-free networks, and analytically for the
exponential trees. Our analytical results are formulated in terms of iterative
equations and they are exact also for small trees.

\section{Numerical calculations and results}

\begin{figure}[tbp]
\begin{center} 
\includegraphics[width=0.3\textwidth]{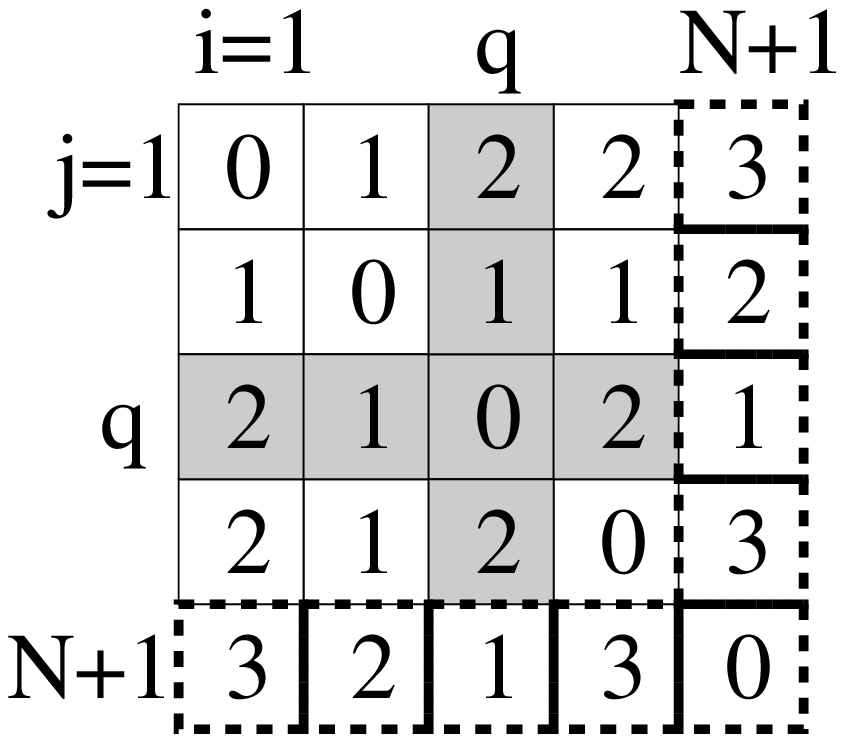} 
\caption{Construction of the distance matrix.}
\label{fig_matrix}
\end{center}
\end{figure}

The numerical algorithm is as follows: an initial tree is composed of two
nodes connected by an edge. Each time when a $(N+1)$-th node is added to a
$q$-th node, the distance $s(N+1,i)$ between the new node and any other (say,
$i$-th one) is $1+s(q,i)$.  The set of distances can be presented as a
symmetric matrix $s(i,j)$. Adding a new node is represented by an enlargement
of the matrix size by one. The last $(N+1)$-th column and row are equal to one
plus the $q$-th column (or row), where $q$ is the index of a node to which the
new node is attached (see Fig.~\ref{fig_matrix}).

\begin{figure}[tbp]
\begin{center} 
\includegraphics[angle=-90,width=0.9\textwidth]{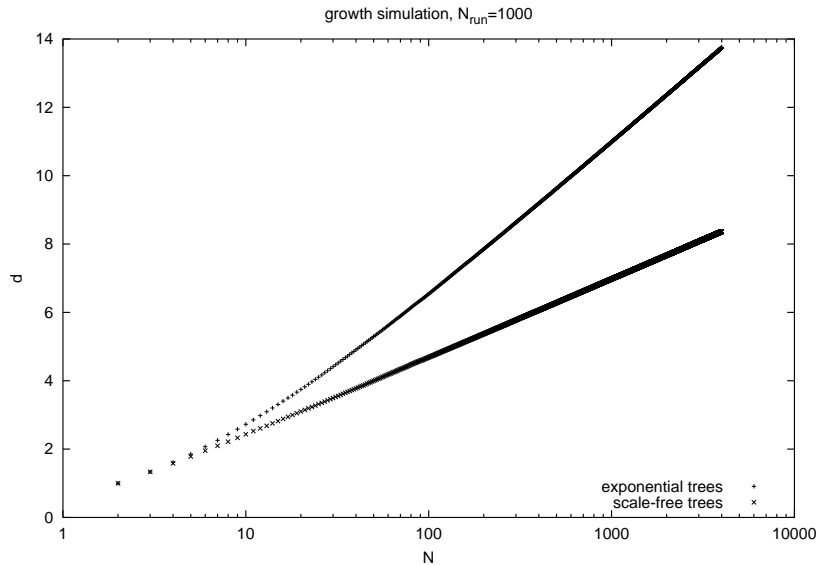} 
\caption{The average distance $d$ for the exponential and the scale-free trees.}
\label{fig_d}
\end{center}
\end{figure}

The mean distance, i.e. the average length $d$, is just
the average over the nondiagonal matrix elements. Obviously, the diagonal
elements are zero. The obtained curve $d(N)$ is an average over $N_{run}=10^3$
trees.

The results on $\langle d\rangle $ (hereafter abbreviated as $d$) for the
exponential trees and for the scale-free trees they are shown in Fig.~\ref{fig_d}.

For large $N$, the obtained curves can be well approximated as linear with
$\ln(N)$. Namely we get $d(N)=1.98\ln(N)-2.69$ for the exponential trees and
$d(N)=1.00\ln(N)+0.04$ for the scale-free trees. Preliminary numerical results
on the dispersion for both kind of trees suggest that for large $N$, the
dispersion  $\sigma_d^2(N)=a\ln(N)+b$, where $a$ is the same as for the
average distance $d$, i.e. $2.0$ and $1.0$ respectively for the exponential
and the scale-free trees.       	

\section{Analytical considerations}

During the tree growth, the size of the matrix $s(i,j)$ of distances between
nodes increases. Then, the time evolution of a tree is equivalent to the
evolution of the matrix size $N$. We consider the evolution of the ensemble
average of the distance matrix $s(i,j)$. The mean distance $d(N)$ between
nodes, averaged over all pairs of nodes $(i,j)$ and trees $t_N$ with $N$ nodes can be
written as
\begin{equation}  \label{eq_1} 
d(N)=\frac{1}{N(N-1)}\sum_t P(t_N)\sum_{i,j\ne i}^N s_{i,j}(t_N),
\end{equation} 
where $P(t_N)$ is the probability of a tree $t_N$ of size $N$, and it is
normalized to unity within the set of all possible trees with $N$ nodes.
Although we are interested in $d$, it is easier to work just with the average
matrix element $s(N)$, which includes diagonal elements $s(i,i)$ 
\begin{equation}  \label{eq_2} 
s(N)=\frac{1}{N^2}\sum_t P(t_N)\sum_{i,j=1}^N s_{i,j}(t_N).
\end{equation} 

\begin{figure}[tbp]
\begin{center} 
\includegraphics[angle=-90,width=0.9\textwidth]{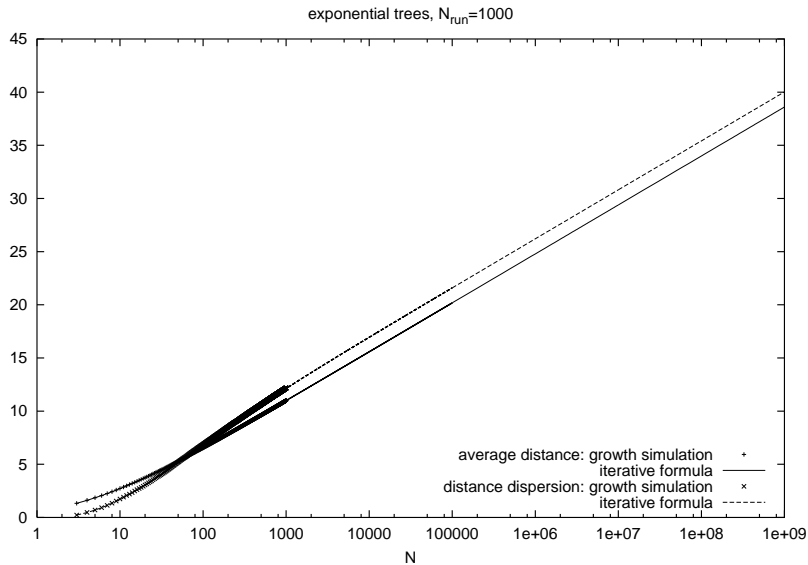} 
\caption{ The average distance $d$ and its dispersion $\sigma_d^2$ for the
exponential trees. Numerical results agree with those obtained from Eqs. (6)
and (8).}
\label{fig_exptre}
\end{center}
\end{figure}
 
The relation $(N-1)d(N)=Ns(N)$ comes from the obvious fact that the diagonal
element $s(i,i)=0$. The evolution of $s(N)$ is given by 
\begin{equation} 
\label{eq_3} 
\begin{split}
&(N+1)^2s(N+1)=\sum_{t}P(t_{N+1})\sum_{i,j=1}^{N+1}s_{i,j}(t_{N+1})=\\
&=\sum_tP(t_{N+1})\sum_{i,j=1}^{N}s_{i,j}(t_{N+1})+2\sum_tP(t_{N+1})
\sum_{j=1}^{N}s_{i,N+1}(t_{N+1}) 
\end{split}
\end{equation} 
because the matrix $s(i,j)$ is symmetrical and $s(N+1,N+1)=0$. In the first
sum of the latter expression, the average over trees with $(N+1)$ nodes is
performed over the sum of matrix elements which depend on the subgraph $t_N$.
That is why in fact $P(t_{N+1})$ is the boundary distribution $P(t_N)$, and the
information on the $(N+1)$-th node is averaged out. In the second sum,
$P(t_{N+1})=P(t_N)P(q|t_N)$, where the latter is the conditional probability of
attaching a new node to a preexisting node indexed by $q$. On the other hand,
we have again $s(i,N+1)=1+s(i,q)$. This is the same rule which has been
helpful in the numerical calculations, described above. Substituting all that
to Eq. (3), we get
\begin{equation} 
\label{eq_4} 
\begin{split}
(N+1)^2s(N+1)=&\sum_tP(t_{N})\sum_{i,j=1}^{N}s_{i,j}(t_N)+2N+\\
&+2\sum_tP(t_N)\sum_{q=1}^NP(q|t_N)\sum_{j=1}^{N}s_{i,q}(t_N).
\end{split}
\end{equation} 

The first term on r.h.s. is nothing but $N^2s(N)$; the problem is with the
last term. For scale-free trees, the probability $P(q|t_N)$ is proportional to
the degree of the node $q$ in the tree $t_N$, i.e. it is equal to $\gamma 
k(q)$, where $\gamma $ is a tree-dependent constant. Substituting it to Eq.
(4), we obtain the relation between the average distance between nodes of
trees with $(N+1)$ nodes and the average distance from a node of degree $k$ to
any other node in trees with $N$ nodes. This relation, even if natural, seems
hard to use further.

On the contrary, for the exponential trees $P(q|t_N)=1/N$. This term
can be extracted out from the sum and we get a compact iterative relation 
\begin{equation}  \label{eq_5} 
(N+1)^2s(N+1)=N^2s(N)+2N+2Ns(N).
\end{equation} 

Substituting $d(N)=Ns(N)/(N-1)$, we get 
\begin{equation}  \label{eq_6} 
d(N+1)=\frac{(N+2)(N-1)}{N(N+1)}d(N)+\frac{2}{N+1}.
\end{equation} 

The same manipulations are possible with the moments of higher order. The
average of the squared matrix element is
\begin{equation}  \label{eq_7} 
e(N)=\frac{1}{N(N-1)}\sum_t P(t_N)\sum_{i,j\ne i}^N s^2_{i,j}(t_N).
\end{equation} 

Applying it to the exponential trees, we get the relation 
\begin{equation}  \label{eq_8} 
e(N+1)=\frac{(N+2)(N-1)}{N(N+1)}e(N)+\frac{4(N-1)}{N(N+1)}d(N)+\frac{2}{N+1}.
\end{equation} 

We start from the only existing tree of $N=2$; obviously $d(2)=1$, $e(2)=1$.
Successive values of $d(N)$, $\sigma_d^2(N)=e(N)-d^2(N)$ are shown in Fig.~\ref{fig_exptre}
together with the respective results of the simulation.

As we see, the agreement is good. For high values of $N$, the
plots are linear with $\ln(N^2)$. For $10^5\le N\le 10^7$, the obtained
coefficients are: $d(N)=2.0\ln(N)-2.84$, $\sigma_d^2(N)=2.0\ln(N)-1.44$.

\section{Conclusions}
For both kinds of trees, the terms of $d$ and $\sigma_d^2$ leading in $N$ are
proportional to $\ln(N)$, with apparently the same coefficients ($2.0$ and
$1.0$ for the exponential trees and the Barab\'asi--Albert scale-free trees,
respectively). This means, that in both cases the term in $\sigma_d^2$
proportional to $\ln^2(N)$ cancels. 

The selecting of subsequent nodes to attach new nodes is equivalent to a
random walk in the space of all trees. At each step, the number of
possibilities --- nodes where a new node can be attached --- increases. In
principle, the number $N_{run}$ of trees used in the averaging procedure should
increase in time to have a good statistics, but such a simulation would be 
time-consuming. If $N_{run}$ is much smaller than the maximal size $N$ of
trees, the averaging is not ergodic and numerical errors can increase.

In our formulation, the weights of particular trees of size $(N+1)$ do depend
on the weights of trees of size $N$ from which they are formed. This dependence
is specific for the process of growing of the trees. In consequence, our
results differ from those obtained within the equilibrium statistical
mechanics, e.g. in \cite{bia}.

\bigskip

\noindent
{\bf Acknowledgements.}
Two of the authors (K.M. and K.K.) are grateful to Prof. Dietrich Stauffer for illuminating discussions.
Part of the calculations were carried out in ACK-CYFRONET-AGH.
The machine time on SGI~2800 is financed by the Polish State Committee for Scientific Research (KBN).


\end{document}